%% file: brst_timelike_liouville.tex
\documentclass[10pt, a4paper]{article}
\pdfoutput=1
\usepackage[utf8]{inputenc}
\usepackage[T1]{fontenc}
\usepackage{lmodern, textcomp}
\usepackage{csquotes}

\usepackage[heightrounded,
	top=4cm, bottom=4cm, left=3.5cm, right=3.5cm,
]{geometry}

\usepackage{eurosym, xspace}
\usepackage[nottoc]{tocbibind}
\usepackage{graphicx, subcaption, float}
\usepackage[multiple]{footmisc}
\usepackage{authblk}
\usepackage{marginnote}
\usepackage{xparse}

\usepackage{fancyhdr}
\usepackage{makecell}
\usepackage{changepage}
\usepackage{multicol}
\usepackage{lastpage}
\usepackage{enumitem}
\usepackage{diagbox}
\usepackage{multirow}
\usepackage{tabularx}
\usepackage{tabu}

\usepackage[usenames, dvipsnames, svgnames,  table]{xcolor}

\usepackage[backend=bibtex8, citestyle=numeric-comp, maxbibnames=99,
			sorting=none, sortcase=false, sortcites=true, giveninits=true]{biblatex}
\input{arxiv.def}

\usepackage{amsmath, amsfonts, amssymb}
\usepackage{mathtools, cancel}
\usepackage{siunitx}
\usepackage{slashed, tensor}

\usepackage{framed, mdframed}
\usepackage[amsmath, hyperref, framed]{ntheorem}

\usepackage[pdftex, breaklinks=true, linktocpage,
	colorlinks=true, urlcolor=blue, linkcolor=blue, citecolor=red,
	pdfauthor={Harold Erbin}
]{hyperref}
\usepackage[all]{hypcap}

\DeclareUnicodeCharacter{00A0}{~}
\DeclareUnicodeCharacter{202F}{~}

\newcommand{\what}[1]{\widehat{#1}}

\newcommand{\email}[1]{\href{mailto:#1}{\nolinkurl{#1}}}
\newcommand{\emailfoot}[1]{\thanks{\email{#1}}}

\newcommand{\doi}[1]{\href{http://dx.doi.org/#1}{\nolinkurl{#1}}}

\newcounter{draftcommentcnt}
\NewDocumentCommand{\draftcomment}{s O{red} m}{%
	\def\margnote{\IfBooleanTF{#1}{\marginnote}{\marginpar}}%
	\stepcounter{draftcommentcnt}%
	\textcolor{#2}{#3}%
	\margnote{\textcolor{#2}{$\Leftarrow$ \arabic{draftcommentcnt}}}%
}

\newcommand{\preprint}[0]{XXX}
\fancypagestyle{preprint}{
	\fancyhf{}

	\fancyhead[R]{\textsf{\preprint}}
}

\fancypagestyle{pageof}{
	\fancyhf{}

	\fancyfoot[C]{Page \thepage\ of \pageref*{LastPage}}
}

\graphicspath{{images/}}

\numberwithin{equation}{section}

\usepackage[sort&compress, english]{cleveref}

\hypersetup{
	pdftitle={BRST cohomology of timelike Liouville},
}

\title{BRST cohomology of timelike Liouville theory}

\author[1]{Teresa Bautista\emailfoot{teresa.bautista@aei.mpg.de}}
\affil[1]{%
	Max Planck Institute for Gravitational Physics (Albert Einstein Institute)
	\protect\\
	Mühlenberg 1, D-14476 Potsdam, Germany}

\author[2]{Harold Erbin\emailfoot{erbin@to.infn.it}}
\affil[2]{%
	Dipartimento di Fisica, Università di Torino
	\protect\\
	\textsc{Infn} Sezione di Torino and Arnold--Regge Center,
	\protect\\
	Via Pietro Giuria 1, I-10125 Torino, Italy
}

\author[3]{Matěj Kudrna\emailfoot{kudrnam@fzu.cz}}
\affil[3]{%
	Institute of Physics of the ASCR, v.v.i. \authorcr
	Na Slovance 2, 182 21 Prague 8, Czech Republic
}

\newcommand{\e}[0]{\mathrm{e}}
\newcommand{\I}[0]{\mathrm{i}}
\newcommand{\N}[0]{\ensuremath{\mathbb{N}}}
\newcommand{\Z}[0]{\ensuremath{\mathbb{Z}}}

\newcommand{\R}[0]{\ensuremath{\mathbb{R}}}
\newcommand{\C}[0]{\ensuremath{\mathbb{C}}}

\newcommand{\dd}[0]{\mathrm{d}}
\newcommand{\pd}[0]{\partial}
\newcommand{\grad}[0]{\nabla}
\newcommand{\mc}[1]{{\mathcal{#1}}}

\renewcommand{\vec}[1]{\ensuremath{\boldsymbol{#1}}}

\newcommand{\ket}[1]{\mathinner{|#1 \rangle}}
\newcommand{\bra}[2][]{\mathinner{{}_{#1}\hspace{-2pt}\langle #2 |}}

\newcommand{\bracket}[2]{\mathinner{\langle #1 | #2 \rangle}}

\newcommand{\abs}[1]{{|#1|}}

\newcommand{\com}[2]{[ #1, #2 ]}
\newcommand{\anticom}[2]{\{ #1, #2 \}}

\newcommand{\group}[1]{\mathrm{#1}}

\begin{document}

\maketitle

\begin{abstract}
We compute the Hermitian sector of the relative BRST cohomology of the spacelike and timelike Liouville theories with generic real central charge $c_L$ in each case, coupled to a spacelike Coulomb gas and a generic transverse CFT. This paper is a companion of~\cite{Bautista:2019:QuantumGravityTimelike}, and its main goal is to completely characterize the cohomology of the timelike theory with $c_L \le 1$ which was defined there.
We also apply our formulas to revisit the BRST cohomology of the spacelike Liouville theory with $c_L > 1$, which includes generalized minimal gravity.
We prove a no-ghost theorem for the Hermitian sector in the timelike theory and for some spacelike models.
\end{abstract}

\newpage

\hrule
\pdfbookmark[1]{\contentsname}{toc}
\tableofcontents
\bigskip
\hrule

\input{sections/introduction}
\input{sections/cft}

\input{sections/brst_derivation}
\input{sections/brst_applications}

\input{sections/discussion}

\section*{Acknowledgements}

We acknowledge useful discussions with Dileep Jatkar, Sylvain Ribault, Raoul Santachiara and Ashoke Sen.
We are particularly grateful to Atish Dabholkar for suggesting us this problem, for ongoing discussions and for collaborations on related topics.
The work of H.E.\ has been conducted under a Carl Friedrich von Siemens Research Fellowship of the Alexander von Humboldt Foundation during part of this project.
H.E.\ is partially supported by the \textsc{Miur Prin} Contract \textsc{2015Mp2cx4} “Non-perturbative Aspects of Gauge Theories and Strings”.

\appendix

\printbibliography[heading=bibintoc]

\end{document}

%% file: sections/introduction.tex
\section{Introduction}
\label{sec:intro}

Liouville theory arises as the universal gravitational action (Wess--Zumino action) of matter described by a conformal field theory (CFT), coupled to two-dimensional gravity in the conformal gauge~\cite{Polyakov:1981:QuantumGeometryBosonic}.
As such, it provides an important toy model for approaching four-dimensional quantum gravity in the conformal limit~\cite{Polchinski:1989:TwodimensionalModelQuantum,Dabholkar:2016:QuantumWeylInvariance, Bautista:2016:QuantumCosmologyTwo}.
Liouville theory is itself a CFT, and when considered as a model of 2d gravity it has central charge $c_L = 26 - c_m$, where $c_m$ is the central charge of the matter CFT.

The Liouville action is characterized by two parameters: the sign of the kinetic term $\epsilon = \pm 1$ (see~\cite{Giribet:2012:TimelikeLiouvilleThreepoint} for an earlier use of this parametrization), and the background charge $Q \in \C$.
The central charge depends on these two parameters as
\begin{equation}
	\label{eq:intro-cL}
	 c_L
		:= 26 - c_m
		:= 1 + 6 \,\epsilon\, Q^2.
\end{equation}
In the CFT language, the parameters $\epsilon$ and $Q$ appear in the OPEs of the $\group{U}(1)$ current and of the energy--momentum tensor.
The theory is called spacelike when $\epsilon = 1$, and timelike when $\epsilon = - 1$.

We are mainly interested in the case where $2d$ gravity is defined by its path integral.
First, it is the most common approach in the literature and it may be more familiar than the CFT approach to most readers.
Second, it directly generalizes to the case where the matter is not a CFT.
In this case, the only known formulation is through the path integral, and the gravitational action is given by the Liouville action plus corrections which are not conformally invariant~\cite{Ferrari:2012:GravitationalActionsTwo, Ferrari:2014:FQHECurvedBackgrounds, Bilal:2014:2DQuantumGravity, Bilal:2017:2DQuantumGravity, Bilal:2017:2DGravitationalMabuchi, deLacroix:2016:MabuchiSpectrumMinisuperspace, deLacroix:2018:MinisuperspaceComputationMabuchi}.
Third, most higher-dimensional theories of quantum gravity are defined from an action.

When defining Liouville gravity from the path integral, it is common to consider the action \eqref{cft:liouville-action} to be real (at least for the Gaussian part), this requires $Q$ to be real.
Having $Q \in \R$ is also more natural when interpreting the Liouville theory as the limit $d \to 2$ of higher-dimensional Einstein--Hilbert gravity~\cite{Kawai:1993:ScalingExponentsQuantum, Kawai:1993:UltravioletStableFixed, Mann:1993:Dto2Limit, Bautista:2016:QuantumCosmologyTwo}.
Indeed, the factor $1 / Q^2$ sitting in front of the action in the semi-classical limit can be identified with an appropriate limit of the $d$-dimensional Newton's constant, which is usually taken as real.
The condition $Q \in \R$ leaves three possible ranges of parameters: spacelike with $c_L \ge 25$ and $Q \ge 2$, spacelike with $c_L \in (1, 25)$ and $Q \in (0, 2)$, and timelike with $c_L \le 1$ and $Q \in \R$.
We will focus on these three regimes when discussing applications of this paper.
However, reality of the action (or, more generally, any condition on the action) is restrictive since there are interesting theories with a complex Lagrangian or which do not have a Lagrangian at all.\footnotemark{}
\footnotetext{%
	Minimal models~\cite{Cappelli:1987:ADEClassificationMinimal} are one such example: it is well-known~\cite[chap.~9]{DiFrancesco:1999:ConformalFieldTheory} that they can be obtained from the Coulomb gas theory with $c \le 1$, which has a complex Lagrangian since $Q \in \I\R$, using Felder's resolution~\cite{Felder:1989:BRSTApproachMinimal, Felder:1989:FreeFieldRepresentation, Bouwknegt:1991:FockSpaceResolutions}.
	Another category of interesting theories with complex Lagrangians are PT-symmetric CFTs~\cite{Bender:2004:ScalarQuantumField, Bender:2005:ExtensionPTSymmetricQuantum, Korff:2008:PTSymmetryNonHermitian, Dorey:2009:PTsymmetricQuantumMechanics, Amoruso:2016:RenormalizationGroupFlows, CastroAlvaredo:2017:IrreversibilityRenormalizationGroup, Ashida:2017:ParitytimesymmetricQuantumCritical}.
}%
Since Liouville theory is a two-dimensional CFT, it can be completely defined using only CFT techniques.
It is within this framework that Liouville theory can be written for $Q \in \C$ without ambiguities.

While the spacelike theory is well understood for $c_L \ge 25$ from different points of view~\cite{Teschner:2001:LiouvilleTheoryRevisited, Nakayama:2004:LiouvilleFieldTheory, Ribault:2014:ConformalFieldTheory, Rhodes:2016:LectureNotesGaussian, Kupiainen:2016:ConstructiveLiouvilleConformal}, it is not the case for the other parameter ranges.
The most interesting case is the timelike theory with $c_L \le 1$, studied extensively in~\cite{Strominger:2002:OpenStringCreation, Gutperle:2003:TimelikeBoundaryLiouville, Strominger:2003:CorrelatorsTimelikeBulk, Schomerus:2003:RollingTachyonsLiouville, Fredenhagen:2003:MinisuperspaceModelsSbranes, Zamolodchikov:2005:ThreepointFunctionMinimal, Kostov:2006:BulkCorrelationFunctions, Kostov:2007:NonRational2DQuantum-1, Kostov:2007:NonRational2DQuantum-2, McElgin:2008:NotesLiouvilleTheory, Harlow:2011:AnalyticContinuationLiouville, Giribet:2012:TimelikeLiouvilleThreepoint}.
Indeed, this theory serves as a toy model for four-dimensional quantum gravity since the kinetic term of the conformal factor is negative definite in $d = 4$~\cite{Gibbons:1977:ActionIntegralsPartition} (see also~\cite{Polchinski:1989:TwodimensionalModelQuantum, Cooper:1991:TwodimensionalQuantumCosmology, Martinec:2014:ModelingQuantumGravity, Dabholkar:2016:QuantumWeylInvariance, Bautista:2016:QuantumCosmologyTwo, Bautista:2019:QuantumGravityTimelike}).
The main obstacle in understanding this regime has been that different quantization procedures (minisuperspace, bootstrap, BRST…) give different spectra.
It was proposed in~\cite{Bautista:2019:QuantumGravityTimelike} to use BRST quantization as the fundamental guiding principle to determine the spectrum, since it encodes the constraints from the diffeomorphism invariance, the gauge symmetry of gravity.

It was recently proven numerically that the spacelike Liouville theory ($\epsilon = +1$) is a consistent CFT for all $c_L \in \C$ as it solves the conformal bootstrap constraints~\cite{Ribault:2015:LiouvilleTheoryCentral, Ribault:2014:ConformalFieldTheory} (see also~\cite{Delfino:2011:ThreepointConnectivityTwodimensional, Picco:2013:ConnectivitiesPottsFortuinKasteleyn, Ikhlef:2016:ThreepointFunctionsc} for a connection to statistical loop models).
The $3$-point function which solves the conformal bootstrap constraints depends only on the value of the central charge: it is given by the DOZZ formula everywhere in the complex $c_L$-plane except for the real strip $c_L \le 1$, where it is instead given by the “$c_L \le 1$ structure constant” (also called timelike DOZZ formula)~\cite{Schomerus:2003:RollingTachyonsLiouville, Kostov:2006:BulkCorrelationFunctions, Kostov:2007:NonRational2DQuantum-1, Kostov:2007:NonRational2DQuantum-2}.
Convergence of the $4$-point function also determines the internal spectrum, defined as the set of states on which the correlation functions factorize.
The internal spectrum is unique for $c_L \in \C$ and made of states with real momenta.
With these ingredients, the $4$-point function is crossing symmetric.

A complete definition of the timelike Liouville theory with $c_L \le 1$ on the sphere has been proposed in~\cite{Bautista:2019:QuantumGravityTimelike}: it provides a spectrum of external states and a definition of the $3$- and $4$-point correlation functions which can be generalized to higher orders.
We provide a brief summary of the results~\cite{Bautista:2019:QuantumGravityTimelike}:
\begin{itemize}
	\item The \textbf{spectrum} is computed in~\cite{Bautista:2019:QuantumGravityTimelike} from the BRST quantization.
	Earlier works on the $c_L\le 1$ Liouville theory~\cite{Strominger:2002:OpenStringCreation, Gutperle:2003:TimelikeBoundaryLiouville, Strominger:2003:CorrelatorsTimelikeBulk, Schomerus:2003:RollingTachyonsLiouville, Fredenhagen:2003:MinisuperspaceModelsSbranes, McElgin:2008:NotesLiouvilleTheory} approached the question using the minisuperspace approximation, but there are subtelties: the Hamiltonian is not Hermitian, the spectrum does not match the one of the conformal bootstrap and displays strange discrete states, and the $3$-point function does not match the $c_L \le 1$ structure constant.
	This motivated the use of the BRST quantization to determine the physical spectrum.
	In~\cite{Bautista:2019:QuantumGravityTimelike}, the BRST cohomology has been computed only when the matter is made of free scalar fields.
	Generalizing this computation to other types of CFT matter is the main objective of this paper.

	In~\cite{Ribault:2015:LiouvilleTheoryCentral}, the spectrum of the theory is defined as the internal states on which the correlation functions factorize.
	However, they are different from the states found in the BRST cohomology.
	For this reason, \cite{Bautista:2019:QuantumGravityTimelike} concludes that this definition is too restrictive and that one must distinguish between the internal spectrum and external or physical spectrum, in the same way that one differentiates between off-shell and on-shell states in QFT.
	As indicated above, the internal spectrum is constrained by the convergence of the $4$-point function, so it cannot be changed arbitrarily.

	\item The \textbf{correlation functions} are described explicitly in~\cite{Bautista:2019:QuantumGravityTimelike} up to four points.
	The $2$-point function is the direct analytic continuation of the spacelike theory one as proposed in~\cite{Harlow:2011:AnalyticContinuationLiouville} since it is analytic in $Q$, well defined for any external complex momentum and matches the saddle-point computations in the semi-classical limit.
	The $3$-point function is given by the $c_L \le 1$ structure constant derived in~\cite{Schomerus:2003:RollingTachyonsLiouville, Zamolodchikov:2005:ThreepointFunctionMinimal, Kostov:2006:BulkCorrelationFunctions, Kostov:2007:NonRational2DQuantum-1, Kostov:2007:NonRational2DQuantum-2}.
	Indeed, the path integral and Coulomb gas computations from~\cite{Harlow:2011:AnalyticContinuationLiouville, Giribet:2012:TimelikeLiouvilleThreepoint} convincingly show that this is the correct $3$-point function for the timelike theory.
	It is also analytic in the external momenta.
	Moreover, solving the full crossing equations shows that the $3$-point function is unique for a given central charge~\cite{Ribault:2014:ConformalFieldTheory, Ribault:2015:LiouvilleTheoryCentral}.

	The timelike $4$-point function is defined by analytic continuation of the spacelike $4$-point function with the same central charge~\cite{Bautista:2019:QuantumGravityTimelike}.
	This analytic continuation is inspired from string field theory~\cite{Pius:2016:CutkoskyRulesSuperstring, Pius:2018:UnitarityBoxDiagram, deLacroix:2017:ClosedSuperstringField, deLacroix:2019:AnalyticityCrossingSymmetry} and generalizes the usual Wick rotation from QFT to the case where the momentum integrand diverges exponentially for large imaginary momenta (real energy).
	Under this procedure, external and internal momenta change differently, which explains the need to distinguish between the external physical spectrum and the internal one used for factorization.
	This provides a solution to the factorization as was called for in the conclusion of~\cite{Harlow:2011:AnalyticContinuationLiouville}.

	Something special happens in the case of the $c_L \le 1$ timelike theory (but would not generalize to other values of the central charge): there is no need to modify the internal states, such that the same $4$-point correlation function as in~\cite{Ribault:2015:LiouvilleTheoryCentral} can be used to define the $c_L \le 1$ timelike theory.
	As a consequence, the difference with~\cite{Ribault:2015:LiouvilleTheoryCentral} for $c_L \le 1$ boils down to the external states.
	In particular, it was proven in~\cite{Ribault:2015:LiouvilleTheoryCentral} that this $4$-point function is crossing symmetric for external states with general complex momenta.
	Hence, this implies that the definition of the timelike theory given in~\cite{Bautista:2019:QuantumGravityTimelike} is crossing symmetric.
\end{itemize}
We refer the reader to the original paper~\cite{Bautista:2019:QuantumGravityTimelike} for more details on the various aspects and on the comparison with previous works.

As indicated above, the main goal of this paper is to completely characterize the BRST cohomology of $c_L \le 1 $ timelike Liouville, thus generalizing the computation in~\cite{Bautista:2019:QuantumGravityTimelike} done solely for a free scalar matter theory.
To this aim, we study the BRST cohomology of Liouville theory for generic values of the parameters $\epsilon$ and $Q$, coupled to a spacelike Coulomb gas and a generic transverse CFT.
This also generalizes previous studies for the spacelike Liouville theories with $c_L \ge 25$~\cite{Bouwknegt:1992:BRSTAnalysisPhysical} (see also~\cite{Bachas:1990:FiniteNumberStates, Mukherji:1991:NullVectorsExtra, Mukhi:1991:ExtraStatesC1, Lian:1991:2DGravityC1, Lian:1991:NewSelectionRules, Chair:1992:SO2CInvariantRing, Imbimbo:1992:ConstructionPhysicalStates, Witten:1992:GroundRingTwo, Kutasov:1992:GroundRingsTheir, Witten:1992:AlgebraicStructuresDifferential, Seiberg:2004:BranesRingsMatrix}) and $c_L \in (1, 25)$~\cite{Bilal:1992:RemarksBRSTcohomologycM}.

The BRST cohomology is computed by generalizing the derivation of Bouwknegt, McCarthy and Pilch~\cite{Bouwknegt:1992:BRSTAnalysisPhysical}, where the Liouville theory is represented by a Coulomb gas.
Indeed, we generalize all formulas to depend on $\epsilon$ and $Q$, so that they are applicable for any Liouville range.
As a consistency check, we rederive the results from~\cite{Bouwknegt:1992:BRSTAnalysisPhysical} for the spacelike Liouville theory ($c_L \ge 25$).
Imposing that the BRST charge is Hermitian truncates the BRST cohomology to a subset: for most cases, the states in this sector are ghost-free.
This proves the \emph{no-ghost theorem} in full generality for the Hermitian timelike Liouville theory, and for some of the spacelike Liouville cases.
Finally, we provide an argument to match the cohomology of the Liouville and Coulomb gas when the Fock space of the latter contains no degenerate states for Hermitian momenta.\footnotemark{}
\footnotetext{%
	Subtleties related to the cosmological constant in the general case and its possible effect on the Hermiticity condition is beyond the scope of this paper; the reader is referred to~\cite{Bautista:2019:QuantumGravityTimelike} for some discussions.
}%
To the best of our knowledge, the connection between Hermiticity, no-ghost theorem and the cosmological constant is a new observation.
Moreover, this generic derivation of the cohomology -- while following directly from known methods -- has never been presented with a unified notation.

The BRST analysis presented in this paper shows that \emph{there are} spectra which distinguish between $\epsilon = \pm 1$ even for identical values of $c_L$, and contrary to the bootstrap.\footnotemark{}
\footnotetext{%
	The central charges of two theories with parameters $(\epsilon = 1, Q)$ and $(\epsilon = - 1, \I Q)$ are equal, see \eqref{eq:intro-cL}.
}%
Moreover, the spectrum given by the conformal bootstrap for $c_L \le 1$ is not compatible with Hermitian Virasoro operators, which also means that the BRST charge is not Hermitian.
On the other hand, the Hermitian sector of the BRST cohomology is not empty for its timelike counterpart.
These differences justify the importance of keeping track of both parameters $\epsilon$ and $Q$.

We also revisit the Liouville theory with $c_L \in (1, 25)$ studied in~\cite{Bilal:1992:RemarksBRSTcohomologycM} by using a different Hermiticity condition.
Finally, it should be possible to consider generalized minimal models~\cite{Zamolodchikov:2005:ThreepointFunctionMinimal} for the matter sector by generalizing Felder's construction~\cite{Felder:1989:BRSTApproachMinimal, Felder:1989:FreeFieldRepresentation, Bouwknegt:1991:FockSpaceResolutions}.
This would provide $2d$ gravity models for all central charges $c_L \ge 25$.
However, the details of generalizing Felder's resolution to generalized minimal models is outside the scope of this paper and left for a future study.

The outline of the paper is as follows.
\Cref{sec:cft} provides a summary of the CFT systems needed in the analysis.
In \Cref{sec:brst_derivation}, we present a general derivation of the BRST cohomology for a CFT with (at least) two Coulomb gases, generalizing known results.
The formulas are then applied to Liouville theory in different regimes in \Cref{sec:brst_applications}, and, in particular, to the timelike theory in \Cref{sec:brst-timelike}.
Finally, we discuss our results in \Cref{sec:discussion}.

%% file: sections/cft.tex
\section{Conformal field theories}
\label{sec:cft}

In this section, we set the notations and gather the relevant expressions for the CFTs involved in this paper: the Coulomb gas, the Liouville theory, and the $bc$ ghosts.
The main goal of this section is to derive unified expressions -- which do not appear comprehensively elsewhere to our knowledge -- which apply to both the spacelike and timelike theories (positive- and negative-definite kinetic terms) by introducing a parameter $\epsilon = \pm 1$, similarly to~\cite{Giribet:2012:TimelikeLiouvilleThreepoint}.
Normal ordering is implicit and we consider only the holomorphic sector in all the paper.

\subsection{Coulomb gas}
\label{sec:cft:cg}

The Coulomb gas CFT~\cite{DiFrancesco:1999:ConformalFieldTheory} consists of a free scalar field $X$ in the presence of a background charge $Q$
\begin{equation}
	\label{Coulomb_gas_action}
	S = \frac{\epsilon}{4\pi} \int \dd^2 \sigma \sqrt{g}\,
		\left(g^{\mu\nu} \pd_\mu X \pd_\nu X + Q R X \right)
\end{equation}
where $R$ is the Ricci scalar.
The field $X$ can be spacelike or timelike\footnotemark{} depending on the sign of the kinetic term:
\footnotetext{%
	This terminology refers to the signature of the corresponding target space.
}%
\begin{equation}
	\epsilon =
		\begin{cases}
			+ 1 & \text{spacelike},
			\\
			- 1 & \text{timelike},
		\end{cases}
		\qquad
		\sqrt{-1} = \I.
\end{equation}
The main advantage of using this parametrization is that it is not necessary to introduce different sets of notation for the spacelike and timelike cases as it is standard in the Liouville literature.
This also allows to present a unified computation of the BRST cohomology for both cases.
Finally, there is no risk to make sign mistakes in complex conjugation by trying to obtain formulas for one theory from the other by analytic continuation.

The charge $Q$ is parametrized in terms of another parameter $b$ as
\begin{equation}
	\label{cft:eq:Q-b}
	Q = \frac{1}{b} + \epsilon b\,,
\end{equation}
which is defined such that the conformal weight \eqref{cft:eq:vertex-dim} of the vertex operator $V_b = \e^{2 b X}$ is $h_b = 1$.

The energy--momentum tensor $T := T_{zz}$ on flat space reads
\begin{equation}
	T = - \epsilon \,(\pd X)^2 + \epsilon\, Q\, \pd^2 X,
\end{equation}
and the associated central charge is
\begin{equation}
	\label{cft:eq:central-charge}
	c = 1 + 6\, \epsilon \,Q^2 .
\end{equation}
The real values of the central charge in terms of $Q$ and $\epsilon$ are summarised in \Cref{tab:regimes}.
The action \eqref{Coulomb_gas_action} changes by a constant term under a constant shift of $X$.
This leads to the conserved current
\begin{equation}
	J = \I \epsilon\, \pd X,
\end{equation}
which is anomalous at the quantum level if $Q \neq 0$.
Correlation functions have been computed in~\cite{Dotsenko:1985:FourpointCorrelationFunctions, Dotsenko:1985:OperatorAlgebraTwodimensional}.

\paragraph{Mode expansions.}

The Fourier expansion of the Coulomb gas field is
\begin{equation}
	X
		= \frac{x}{2}
			- \epsilon \,\alpha\ln z
			+ \frac{\I}{\sqrt{2}} \sum_{n \neq 0} \frac{\alpha_n}{n}\, z^{-n}.
\end{equation}
where we use the rescaled zero-mode $\alpha := \I \epsilon \, \alpha_0 / \sqrt{2}$.
The zero-mode $\alpha$ is related to the charge of the conserved current as
\begin{equation}\label{charge_shift_field}
		\alpha = \frac{1}{4\pi} \oint \dd z\,  J.
\end{equation}
It is hence interpreted as the momentum on the plane and, because of the current's anomaly, it is related to the momentum $P$ on the cylinder as
\begin{equation}
	\label{cft:eq:cg:relation-alpha-p}
	\alpha = \epsilon \, \frac{Q}{2} + \I P.
\end{equation}

The Virasoro operators are
\begin{equation}
	\label{eq:cg-Ln}
L_m
		= \frac{\epsilon}{2} \sum_{n \neq 0} \alpha_n \alpha_{m-n}
			+ \frac{\I}{\sqrt{2}} \big(\epsilon \,Q (m + 1) - \alpha \big) \alpha_m \,.
\end{equation}
The expression of the zero-mode can be simplified to
\begin{equation}
	\label{Virasoro-zero-mode}
	L_0
		= N + \alpha \left( Q - \epsilon\, \alpha \right)
		= N +  \epsilon\left(\frac{Q^2}{4} + P^2 \right),
\end{equation}
where the level operator $N$ is defined in terms of the number operators $N_n$ at level $n > 0$ as
\begin{equation}
	\label{cft:def:level-and-number-ops}
	N = \sum_{n > 0} n\, N_n \,,
	\qquad
	N_n = \frac{\epsilon}{n}\, \alpha_{-n} \alpha_n \,.
\end{equation}
The canonical commutation relations are
\begin{equation}
	\com{x}{\alpha}
		= - \frac{\epsilon}{2} \, ,
	\qquad
	\com{x}{P}
		= \I \frac{\epsilon}{2} \, ,
	\qquad
	\com{\alpha_m}{\alpha_n}
		= \epsilon\, m\, \delta_{m+n,0}.
\end{equation}
The commutator of the modes with the Virasoro operators is
\begin{equation}
	\com{L_m}{\alpha_n}
		= - n\, \alpha_{m+n} + \I \frac{Q}{\sqrt{2}}\, m (m + 1) \, \delta_{m+n,0}.
\end{equation}
The commutators of the creation modes ($n > 0$) with the number operators are
\begin{equation}
	\com{N_m}{\alpha_{-n}} = \alpha_{-m} \delta_{m,n}.
\end{equation}

\paragraph{Fock space.}

The operator $\pd X$ is primary with conformal weight $h = 1$ only if $Q = 0$.
For any $Q$, the vertex operators
\begin{equation}
	\label{cft:eq:state}
	V_a = \e^{2 a X}
\end{equation}
are primaries and eigenstates of the zero-mode $\alpha$ with $\alpha= a \in \C$.
Their conformal weight is
\begin{equation}
	\label{cft:eq:vertex-dim}
	h_a = a \, (Q - \epsilon\, a).
\end{equation}
According to the anomalous shift \eqref{cft:eq:cg:relation-alpha-p}, they correspond to the operators
\begin{equation}
	V_p = \e^{2 \I p X},
	\qquad
	h_p = \frac{\epsilon\, Q^2}{4} + \epsilon\, p^2
\end{equation}
on the cylinder, with eigenvalue $P = p$ such that the eigenvalues are related by
\begin{equation}
	\label{cft:eq:vertex-relation-a-p}
	a = \frac{\epsilon Q}{2} + \I p.
\end{equation}
Henceforth we will use $V_p$ or $V_a$ indistinguishably.
From the expression of the conformal weights, it follows that the operators $V_a$ and $V_{\epsilon Q-a}$, or equivalently $V_p$ and $V_{-p}$, have the same weight:
\begin{equation}
	h_a = h_{\epsilon Q - a},
	\qquad
	h_p = h_{-p}.
\end{equation}

Note that $h_a, h_p \in \R$ only if $Q, p \in \R \cup \I \R$, which also implies $c \in \R$.
Hence, it makes sense to restrict ourselves to these values of the parameters (however, it is possible to define CFTs for complex values of the central charge and conformal weights~\cite{Ribault:2015:LiouvilleTheoryCentral}).

A set of Fock vacua $\ket{p}$ are obtained by acting with the vertex operators on the $\group{SL}(2, \C)$ vacuum $\ket{0}$
\begin{equation}
	\label{cft:def:vacua}
	\ket{p} = V_p \ket{0}.
\end{equation}
The Fock space $\mc F(p)$ of the theory is generated by all the states obtained by applying creation operators $\alpha_{-n}$ with $n > 0$ on the vacuum
\begin{equation}
	\ket{\psi} = \prod_{n \ge 1} \frac{(\alpha_{-n})^{N_n}}{\sqrt{n^{N_n} N_n!}} \ket{p},
	\qquad
	N_n \in \N.
\end{equation}

The momenta and conformal weights of degenerate states are given by:
\begin{equation}
	\label{eq:degen-momenta}
	p_{r,s}
		= \I \left( \frac{r b}{2} + \frac{s}{2 b} \right),
	\qquad
	a_{r,s}
		= \frac{\epsilon Q}{2} + \I p_{r,s},
	\qquad
	h_{r,s}
		= \frac{\epsilon Q^2}{4} - \epsilon \left( \frac{r b}{2} + \frac{s}{2 b} \right)^2,
\end{equation}
for all $r, s \in \N$.

\paragraph{Hermiticity conditions.}

The Virasoro modes are Hermitian $(L_n)^\dagger = L_{-n}$ if the Coulomb gas modes satisfy (signs are correlated across)~\cite{Itoh:1990:BRSTQuantizationPolyakovs, Frau:1990:OperatorFormalismFree, Kiritsis:2007:StringTheoryNutshell, Ribault:2014:ConformalFieldTheory}
\begin{equation}
	\label{cft:eq:hermiticity-coulomb}
	Q^* = \pm Q,
	\qquad
	\alpha_n^\dagger = \pm \alpha_{-n},
	\qquad
	\alpha^\dagger = \pm (\epsilon Q - \alpha),
	\qquad
	P^\dagger = \pm P,
	\qquad
	x^\dagger = \pm x.
\end{equation}
This implies that $Q \in \R \cup \I \R$ and $p \in \R \cup \I \R$.
The first condition gives $c \in \R$ while both together give $h_p \in \R$.
The Hermiticity condition is chosen such that $Q \in \R \to 0$ reproduces the standard results for the free scalar CFT (in particular, that its momentum is Hermitian).
Further, given that $\alpha = a$ for the vertex operators, it follows that $a^* = \pm (\epsilon Q - a)$.

Comparing the expressions of the degenerate momenta \eqref{eq:degen-momenta} and of the Hermiticity conditions \eqref{cft:eq:hermiticity-coulomb}, we find that they are incompatible for both $\epsilon = \pm 1$ and $c_L \notin (1, 25)$.
Indeed, outside this range, $b$ and $Q$ are both either real or imaginary.
Thus, \eqref{cft:eq:hermiticity-coulomb} implies that $p$ and $b$ must be both either real or imaginary, but the presence of $\I$ in \eqref{eq:degen-momenta} makes this impossible.

\subsection{Liouville theory}
\label{sec:cft:liouville}

Liouville theory corresponds to a Coulomb gas deformed by an exponential interaction representing the coupling to the cosmological constant $\mu$
\begin{equation}
	\label{cft:liouville-action}
	S_L = \frac{\epsilon}{4\pi} \int \dd^2 \sigma \sqrt{g}\,
		\left( g^{\mu\nu} \pd_\mu \phi \pd_\nu \phi + Q R \phi + 4\pi\epsilon \, \mu\, \e^{2 b \phi} \right).
\end{equation}
Thanks to the relation \eqref{cft:eq:Q-b} between $b$ and $Q$, this is a marginal deformation.

\paragraph{Fock space.}

The presence of the exponential potential in the Liouville action entails a time-dependence of the field Fourier modes.
For this reason, the  mode expansion of the Liouville field $\phi$ is complicated.
However, the conformal bootstrap program (see~\cite{Teschner:2001:LiouvilleTheoryRevisited, Ribault:2014:ConformalFieldTheory} for reviews) shows that the theory can be completely defined in terms of the current $J$ and of the vertex operators $V_a$, which satisfy the same properties as the Coulomb gas ones.
This implies that the central charge \eqref{cft:eq:central-charge}, the definition of $Q$ \eqref{cft:eq:Q-b} in terms of $b$, the relation between the plane and cylinder momenta \eqref{cft:eq:cg:relation-alpha-p} and the conformal weights \eqref{cft:eq:vertex-dim} are identical.
The intuition is that, in the regime of small parameter $b$, the exponential wall is only relevant for high values of the field.
For most of the field range, the theory is effectively a Coulomb gas.
One can then understand the effect of the cosmological constant as a reflecting wall which mixes the positive- and negative-frequency modes.
As a consequence, Liouville vertex operators $\mc V_p$ are linear combinations of $V_{\pm p}$ such that
\begin{equation}\label{reflected-op}
	\mc V_{p}
		:= V_p + R(p) V_{-p}
		= R(p) \mc V_{-p},
\end{equation}
where $R(p)$ is the reflection coefficient and satisfies $R(p) R(-p) = 1$.
It can be shown~\cite{Ribault:2014:ConformalFieldTheory} that $R(p)$ is proportional to the cosmological constant $\mu$ and vanishes when the latter is set to zero.

\paragraph{Liouville regimes.}

As for the Coulomb gas, the difference between spacelike and timelike is only determined by the sign of the kinetic term, i.e.\ $\epsilon$ (or of the current--current OPE in the CFT language).\footnotemark{}
\footnotetext{%
	In~\cite{Ribault:2015:LiouvilleTheoryCentral}, the range of the spectrum is used to define the theory as spacelike or timelike.
	However, this is a less primitive concept (compared to the OPE) and, as emphasized in~\cite{Bautista:2019:QuantumGravityTimelike}, it is possible to give several definitions for the spectrum of a theory.
}%
Together with the value or range of $Q$, both parameters completely define the Liouville theory under consideration, giving rise to different regimes.

The most studied and well-known regimes are the ones where the background charge $Q$ is real.
This implies that the central charge is real, and that the (quadratic part of the) action is also real, a property which has been motivated in the introduction.
Like for the Coulomb gas, spacelike Liouville, $\epsilon = 1$, corresponds then to a central charge $c_L \ge 1$.
For the exponential term in the action to be real, it is further necessary for $b$ to be real.
This implies $Q \ge 2$, that is $c_L \ge 25$, which is the regime typically referred to as “\emph{the} spacelike Liouville theory”.
Giving up on the reality of the exponential gives the range $b \in \I \R$ or $c_L \in (1, 25)$.

The regime typically referred to as “\emph{the} timelike Liouville theory”  corresponds to a timelike kinetic term $\epsilon = - 1$ and real $Q$, hence with $c_L \le 1$.
It can also be obtained after an analytic continuation of the field and all parameters (except for the cosmological constant) in the action.

We emphasize that another two regimes compatible with a real central charge exist: that of spacelike Liouville with $Q \in \I \R$ hence $c_L \le 1$ (considered in~\cite{Ribault:2015:LiouvilleTheoryCentral}), and that of timelike Liouville also with $Q \in \I\R$ and hence $c_L > 1$.
All different regimes compatible with real central charge are summarized in \Cref{tab:regimes}.

Finally, the most general regime follows from letting the background charge, or equivalently the central charge, to be complex. The description of $2d$ gravity coupled to conformal matter only requires the total central charge cancellation $c_m + c_L = 26$ and, as such, the regimes with $Q$ imaginary or complex should also be regarded as plausible gravitational models.
Such cases can be important when one is interested in non-Lagrangian theories, in which case the reality of the action is not a relevant feature.
One aim of this paper is precisely to consider all possible regimes by introducing the distinguishing notation $(\epsilon, Q)$, and elucidate the differences between the different regimes, especially within each pair with the same central charge.

Depending on the central charge, two different $3$-point functions are compatible with the degenerate crossing equations of the Liouville theory: the first, for $c_L \notin (-\infty, 1)$, is given by the DOZZ formula~\cite{Dorn:1994:TwoThreepointFunctions, Zamolodchikov:1996:StructureConstantsConformal, Ribault:2015:LiouvilleTheoryCentral}, while the second is valid for $c_L \notin (25, \infty)$~\cite{Zamolodchikov:2005:ThreepointFunctionMinimal, Kostov:2006:BulkCorrelationFunctions, Kostov:2007:NonRational2DQuantum-1, Kostov:2007:NonRational2DQuantum-2, Ribault:2015:LiouvilleTheoryCentral}.
However, the latter range gets restricted to $c_L \in (-\infty, 1]$ when considering the full set of crossing equations~\cite{Ribault:2015:LiouvilleTheoryCentral}.
Note that these two ranges are specified by the value of the central charge, regardless of the value of $\epsilon = \pm 1$ because the bootstrap is insensitive to the sign of the current--current OPE.
Hence, the choice of the $3$-point function is uniquely fixed by the central charge, not by the regime.

Finally, the conformal bootstrap selects a specific spectrum, the internal spectrum, for the OPE to ensure that the $4$-point function is well defined (convergence of the integration over the internal states given by the OPE).
It is characterized by $p \in \R$ for $\epsilon = 1$, $p \in \I \R$ for $\epsilon = - 1$, but since the associated operators have the same conformal weights (and the same lower bound $(c_L - 1) / 24$) for identical central charges, they can be identified (remember that the definition of $p$ is $\epsilon$-dependent).
However, this does not prevent to consider different spectra -- which may be required by other consistency conditions  -- if the theory can be consistently defined by analytic continuation~\cite{Bautista:2019:QuantumGravityTimelike}.

\begin{table}[ht]
\begin{center}
\begin{tabu}{c | [1pt] c | c | c}
	\multirow{2}{*}{$c$}
		& \multicolumn{2}{c |}{$Q \in \mathbb{R}$}
		& \multirow{2}{*}{$Q \in \I \mathbb{R}$}
	\\
	\cline{2-3}
	\multirow{2}{*}{}
		& $Q \in [0,2)$
		& $Q \in [2,\infty)$
		& \multirow{2}{*}{}
	\\
	\tabucline[1pt]{-}
	$\text{spacelike:} \,\,\epsilon= +1 $
		& $c\in [1,25)$
		& $c \ge 25$
		& $c < 1$
	\\
	\hline
	$\text{timelike:} \,\,\,\,\epsilon= -1 $
		& \multicolumn{2}{c |}{$c < 1$}
		& $c > 1$
\end{tabu}
\caption{%
	Range of real values of the Liouville (or Coulomb gas) central charge depending on the parameters $Q$ and $\epsilon$.
	For the spacelike case $\epsilon = 1$, the two different ranges $Q \in [0, 2)$ and $Q \ge 2$ correspond to $b \in \e^{\I\R}$ and $b \in \R$ respectively.
	What is typically known as the spacelike Liouville theory corresponds to the regime $\epsilon=1$ and $Q \ge 2$, and the range typically known as the timelike Liouville theory corresponds to $\epsilon=-1$ and $ Q\in\R$.
	}
\label{tab:regimes}
\end{center}
\end{table}

\paragraph{Analytic continuation.}

The theories $(\epsilon = 1, Q \in \R)$ and $(\epsilon = - 1, Q \in \I \R)$ are related by the following analytic continuation:
\begin{equation}
	\phi = \I \chi,
	\qquad
	Q = \I q,
	\qquad
	b = - \I \beta,
	\qquad
	a = - \I \alpha,
	\qquad
	p = - \I E.
\end{equation}
Indeed, starting from \eqref{cft:liouville-action} with $\epsilon=1$, this analytic continuation yields the timelike Liouville action:
\begin{equation}
	\label{cft:timelike-liouville-action}
	S_{tL} = \frac{1}{4\pi} \int \dd^2 \sigma \sqrt{g}\,
		\left( -g^{\mu\nu} \pd_\mu \chi \pd_\nu \chi - q R \chi + 4\pi \, \mu\, \e^{2 \beta \chi} \right).
\end{equation}
This is usually how one gets “\emph{the} timelike Liouville theory” from “\emph{the} spacelike Liouville theory” at the classical level, as presented in the common literature.
However, this is only a simple way to translate certain classical expressions from one theory to the other (in the usual case, both $Q$ and $q$ are taken to be real).

Most quantities (like $3$-point correlation functions and higher), though, are not analytic in the central charge, such that this analytic continuation cannot be used to derive the properties of the timelike theory from those of the spacelike one~\cite{Schomerus:2003:RollingTachyonsLiouville, Fredenhagen:2003:MinisuperspaceModelsSbranes, Zamolodchikov:2005:ThreepointFunctionMinimal, McElgin:2008:NotesLiouvilleTheory, Harlow:2011:AnalyticContinuationLiouville, Ribault:2015:LiouvilleTheoryCentral, Ribault:2014:ConformalFieldTheory}.
Following~\cite{Bautista:2019:QuantumGravityTimelike}, another proposal is to define the timelike theory at a given $c_L$ as the analytic continuation of the spacelike theory \emph{with the same} $c_L$.
Then, this analytic continuation is not performed on the complex plane of the central charge, but rather on the energies of external and internal states through a generalized Wick rotation~\cite{Pius:2016:CutkoskyRulesSuperstring, Pius:2018:UnitarityBoxDiagram, deLacroix:2017:ClosedSuperstringField, deLacroix:2019:AnalyticityCrossingSymmetry}.
It was shown in~\cite{Bautista:2019:QuantumGravityTimelike} how such an analytic continuation preserves the finiteness and crossing-symmetry of 4-point correlators, thus avoiding the singularities encountered when analytically continuing on the central charge plane.

\subsection{Ghosts}

In two-dimensional gravity, the gauge fixing of the metric to the conformal gauge introduces $b$ and $c$ ghosts with action
\begin{equation}
	S_{\text{gh}}
		= \frac{1}{4\pi} \int \dd^2 \sigma \sqrt{g}\,
			b_{\mu\nu} \big( \grad^\mu c^\nu + \grad^\nu c^\mu - g^{\mu\nu} \grad_\rho c^\rho \big).
\end{equation}
The energy--momentum tensor on the plane reads
\begin{equation}
		T^{\text{gh}}
		= - \pd(b c) - b \pd c\,
\end{equation}
from which it is straightforward to compute the central charge and conformal weights of the ghosts
\begin{equation}
	c_{\text{gh}} = - 26,
	\qquad
	h_b = 2,
	\qquad
	h_c = - 1.
\end{equation}

The ghost action is invariant under an anomalous $\group{U}(1)$ global symmetry with current $j = - b\, c$.
The associated charge is called the ghost number $N^{\text{gh}}$ and is normalised such that
\begin{equation}
	N^{\text{gh}}(b) = - 1,
	\qquad
	N^{\text{gh}}(c) = 1
\end{equation}
on the plane.

\paragraph{Mode expansions.}

The mode expansions of the ghosts are
\begin{equation}
	b(z) = \sum_n b_n\, z^{- n - 2}, \qquad
	c(z) = \sum_n c_n\, z^{- n + 1}.
\end{equation}
The Virasoro operators are
\begin{equation}
	L^{\text{gh}}_m = \sum_n (m - n)\, b_{m+n} c_{-n} \,,
	\qquad
	L^{\text{gh}}_0 = N^b + N^c - 1,
\end{equation}
where the zero-mode is written in terms of the ghost level and number operators
\begin{equation}
	N^b = \sum_{n > 0} n\, N_n^b \,,
	\quad
	N_n^b = b_{-n} c_n \,,
	\qquad
	N^c = \sum_{n > 0} n\, N_n^c \,,
	\quad
	N_n^c = c_{-n} b_n \,.
\end{equation}
The anticommutation relations between the ghosts are
\begin{equation}
	\anticom{b_m}{c_n} = \delta_{m+n,0} \,,
\end{equation}
which imply that $b_n$ and $c_n$ with $n > 0$ are respectively annihilation operators for $c_{-n}$ and $b_{-n}$.
The commutation relations with the Virasoro and number operators are
\begin{subequations}
\begin{gather}
	\com{L^{gh}_m}{b_n} = (m - n)\, b_{m+n},
	\qquad
	\com{L^{gh}_m}{c_n} = - (2 m + n)\, c_{m+n},
	\\
	\com{N_m^b}{b_{-n}} = b_{-m} \delta_{m,n},
	\qquad
	\com{N_m^c}{c_{-n}} = c_{-m} \delta_{m,n}.
\end{gather}
\end{subequations}

\paragraph{Fock space.}

The $\group{SL}(2, \C)$-invariant vacuum $\ket{0}$ is defined by
\begin{equation}
	\forall n \ge -1:
		\quad
		b_n \ket{0} = 0,
	\qquad
	\forall n \ge 2:
		\quad
		c_n \ket{0} = 0.
\end{equation}
However, there exists a $2$-fold degenerate state with a lower energy since $\ket{0}$ is not annihilated by $c_1$.
The degeneracy arises because $b_0$ and $c_0$ commute with the Hamiltonian.
The two ground states are given by
\begin{equation}
	\label{eq:ghost-vacua}
	\ket{\downarrow} = c_1 \ket{0},
	\qquad
	\ket{\uparrow} = c_0 c_1 \ket{0}.
\end{equation}
They are annihilated by all positive frequency modes $b_n$, $c_n$ with $n > 0$, and are related as
\begin{equation}
		b_0 \ket{\downarrow} = 0,
		\qquad
		c_0 \ket{\downarrow} = \ket{\uparrow},
		\qquad
		c_0 \ket{\uparrow} = 0,
		\qquad
		b_0 \ket{\uparrow} = \ket{\downarrow}.
\end{equation}
The two ghost ground states have a vanishing norm and their inner product is normalised to one
\begin{equation}
	\bracket{\downarrow}{\downarrow}
		= \bracket{\uparrow}{\uparrow}
		= 0,
	\qquad
	\bracket{\downarrow}{\uparrow}
		= \bra{0} c_{-1} c_0 c_1 \ket{0}
		= 1.
\end{equation}

By analogy with the critical string, we take $\ket{\downarrow}$ to be the physical vacuum and we use it to build the Fock space $\mc F_{\text{gh}}$ by acting with the creation and annihilation operators
\begin{equation}
	\ket{\psi}
		= c_0^{N_0^c} \prod_{n \ge 1} (b_{-n})^{N_n^b} (c_{-n})^{N_n^c}\, \ket{\downarrow},
	\qquad
	N_n^b, N_n^c = 0, 1.
\end{equation}

\paragraph{Hermiticity conditions.}

The Virasoro modes are Hermitician $(L^{\text{gh}}_n)^\dagger = L^{\text{gh}}_{-n}$ if
\begin{equation}
	b_n^\dagger = b_{-n} \,,
	\qquad
	c_n^\dagger = c_{-n} \,.
\end{equation}

%% file: sections/brst_derivation.tex
\section{BRST cohomology: general derivation}
\label{sec:brst_derivation}

Consider a general CFT with a Coulomb gas scalar field $X$ and a Liouville field $\phi$, with background charges $Q_X, Q_\phi$, and a generic transverse CFT.
In this section, we derive the relative BRST cohomology and the conditions under which the BRST charge is Hermitian.
Following the standard literature, the Liouville field mode expansions are taken to be those of a Coulomb gas.\footnote{We do not discuss in this paper possible subtleties coming from the presence of the cosmological constant and the exponential potential wall.
See~\cite{Bautista:2019:QuantumGravityTimelike} for a discussion.}
The field $X$ is taken to be spacelike, while the field $\phi$ can be spacelike or timelike, i.e.\ formulas depend on $\epsilon_\phi$.
We consider all values of $Q_\phi \in \R, \I \R$, for which the central charge of $\phi$ is real and for which the BRST operator can be Hermitian.
This generalizes to all regimes $(\epsilon, Q)$ the results from~\cite{Bouwknegt:1992:BRSTAnalysisPhysical, Bilal:1992:RemarksBRSTcohomologycM} for spacelike Liouville, and those of~\cite{Bautista:2019:QuantumGravityTimelike} to a general transverse CFT.

\subsection{Setup}

The quantities associated to each scalar are distinguished by an index $\phi$ or $X$, placed as a superscript for the modes, as a subscript otherwise.
For example, the mode operators of the two scalar fields $\phi$ and $X$ are written as $\alpha^\phi_n$ and $ \alpha^X_n$, and the zero modes as $\alpha_\phi$ and $\alpha_X$.
Together, the two scalar fields and the ghosts form the longitudinal sector.
The transverse CFT is unitary and is only characterized by its energy--momentum tensor $T^\perp$ (we do not need to be more precise).
The Hilbert space $\mc H$ of the theory is
\begin{equation}
	\mc H
		= \mc H_\|
			\otimes \mc H_\perp,
	\qquad
	\mc H_\|
		:= \mc F_\phi(p_\phi) \otimes \mc F_X(p_X)
			\otimes \mc F_{\text{gh}},
\end{equation}
where $p_\phi$ and $p_X$ correspond to the momenta of the scalar vacua and $\mc H_\perp$ is the Hilbert space of the transverse CFT.
A basis state of this Hilbert space is given by
\begin{equation}
	\ket{\psi}
		= c_0^{N_0^c} \prod_{m > 0} (\alpha^\phi_{-m})^{N_m^\phi} (\alpha^X_{-m})^{N_m^X}\, (b_{-m})^{N_m^b} (c_{-m})^{N_m^c}\, \ket{p_\phi, p_X, \downarrow}\otimes \ket{\psi_\perp}
\end{equation}
with $\ket{\psi_\perp}$ a state of the transverse CFT, $\ket{p_\phi, p_X, \downarrow}:=\ket{p_\phi} \otimes \ket{p_X} \otimes \ket{\downarrow}$.
The total Virasoro zero-mode operator $L_0$ is
\begin{equation}
	\label{brst:eq:L0-total}
	L_0
		= L^\perp_0 - m^2 - 1 + \what L_0^\|
\end{equation}
where $\what L_0^\|$ is the total longitudinal level operator
\begin{equation}
	\label{brst:eq:level-operator}
	\what L_0^\|
		= N^\phi + N^X + N^b + N^c,
\end{equation}
and the mass $m^2$ corresponds to the vacuum energy of the scalars
\begin{equation}
	\label{brst:eq:mass}
	- m^2
		= a_X (Q_X - a_X) + a_\phi (Q_\phi - \epsilon_\phi a_\phi)
		= \frac{Q_X^2 + \epsilon_\phi Q_\phi^2}{4} + p_X^2 + \epsilon_\phi p_\phi^2.
\end{equation}
The total central charge vanishes as required by gauge invariance of the two-dimensional gravitational theory:
\begin{equation}
	c_{\text{tot}}
		= 2 + 6 (Q_X^2 + \epsilon_\phi Q_\phi^2) + c_\perp - 26
		= 0.
\end{equation}
This leads to a condition on the charges:
\begin{equation}
	\label{brst:eq:square-charges}
	 Q_X^2 + \epsilon_\phi Q_\phi^2
		= 4 - \frac{c_\perp}{6}
\end{equation}
with which the total Hamiltonian can be written in terms of the transverse CFT central charge and the momenta of the two scalars:
\begin{equation}
	\label{brst:eq:L0-total-simp}
	L_0
		= \left( L^\perp_0 - \frac{c_\perp}{24}
			+ \big( p_X^2 + \epsilon_\phi p_\phi^2 \big) \right)
			+ \what L_0 ^\|.
\end{equation}

\paragraph{BRST operator.}

In the following, we use the superscripts “$gh$” and “$m$” to denote the ghost and matter sectors, where the latter includes the two scalars and the transverse CFT.
The BRST current is given by:
\begin{equation}
	j_B(z)
		= c(z) T^m(z) + \frac{1}{2}\, c(z) T^{gh}(z) + \frac{3}{2}\, \pd^2 c(z).
\end{equation}
It is primary only if $c_m = 26$.
The mode expansion of the associated conserved charge $Q_B$ reads
\begin{equation}
	\label{general_BRST_charge}
	Q_B
		= \sum_n c_{n} \left( L^m_{-n} + \frac{1}{2}\, L^{gh}_{-n} \right)
		= \sum_n c_{n} L^m_{-n} + \frac{1}{2} \sum_{m,n} (n - m)\, c_{-m} c_{-n} b_{m+n} - c_0.
\end{equation}
Its ghost number is $N_{\text{gh}}(Q_B) = 1$ and it is nilpotent $Q_B^2 = 0$ if $c_m = 26$.
Importantly, the Virasoro operators are BRST exact:
\begin{equation}
	\label{Virasoro_BRST_ghost}
	L_n = \anticom{Q_B}{b_n},
	\qquad
	\com{Q_B}{L_m} = 0.
\end{equation}

\subsection{Relative cohomology}

Physical states are those states in the Hilbert space which belong to the BRST cohomology $\mc H_{\text{abs}}$, i.e.\ which are $Q_B$-closed but non-exact:
\begin{equation}
	\mc H_{\text{abs}}(Q_B, \mc H)
		= \Big\{ \ket{\psi} \in \mc H \bigm| Q_B \ket{\psi} = 0,
			\nexists \ket{\chi} \in \mc H :
			\ket{\psi} = Q_B \ket{\chi}
			\Big\}.
\end{equation}
The subscript refers to the absolute cohomology, as opposed to the relative cohomology which will be defined shortly.

The general method to construct the absolute cohomology follows~\cite{Bouwknegt:1992:BRSTAnalysisPhysical}.
Other works and reviews include~\cite{Itoh:1990:BRSTQuantizationPolyakovs, Mukhi:1991:ExtraStatesC1, Bouwknegt:1992:BRSTAnalysisPhysicalSugra, Bilal:1992:RemarksBRSTcohomologycM, Itoh:1992:SpectrumTwoDimensionalSuperGravity, Ohta:1992:DiscreteStatesTwoDimensional, Distler:1992:NewDiscreteStates, Polchinski:2005:StringTheory-1}.
The strategy is to find a sequence of isomorphisms between cohomologies of simpler BRST operators.
This is achieved by finding a “contracting homotopy” operator which inverts the BRST operator in some subspace.
Then, one can restrict the BRST operator in the orthogonal subspace to compute the cohomology, because a BRST closed state with a definite eigenvalue of the contracting homotopy operator is necessarily exact.
Restricting to this subspace is what defines the relative cohomology, in which the BRST operator takes a simpler form.
Introducing a light-cone parametrization and iterating the procedure allows to construct the states explicitly.
Finally, one needs to map them to the original space, which is an easy task when the states have no ghosts beyond the one of the vacuum.

A necessary condition for a state $\ket{\psi}$ to be an element of the BRST cohomology is to be on-shell, i.e.\ that its conformal weight vanishes:
\begin{equation}
	L_0 \ket{\psi} = 0.
\end{equation}
Indeed, if $\ket{\psi}$ is closed but not on-shell, $Q_B \ket{\psi} = 0$ and $L_0 \ket{\psi} \neq 0$, then one can use the identity $L_0 = \anticom{Q_B}{b_0}$ to write:
\begin{equation}
	\label{brst:eq:off-shell-psi}
	\ket{\psi}
		= \frac{1}{L_0}\, \anticom{Q_B}{b_0} \ket{\psi}
		= \frac{1}{L_0}\, Q_B \big( b_0 \ket{\psi} \big).
\end{equation}
The state $b_0 \ket{\psi}$ corresponds to another state in the Hilbert space.
As a consequence, $\ket{\psi}$ is exact and does not belong to the cohomology.

It is convenient then to consider the subspace of on-shell states that further satisfy the condition $b_0 = 0$ (Siegel gauge), which we denote by $\mc H_0$
\begin{equation}
	\label{brst:eq:on-shell-fock-space}
	\mc H_0
		= \Big\{ \ket{\psi} \in \mc H \bigm| L_0 \ket{\psi} = 0, b_0 \ket{\psi} = 0 \Big\},
\end{equation}
since the additional condition $b_0 \ket{\psi} = 0$ is sufficient to ensure on-shellness of BRST closed states.
The relative cohomology is defined as the restriction of the absolute cohomology on this subspace:
\begin{equation}
	\mc H_{\text{rel}}(Q_B, \mc H)
		= \Big\{ \ket{\psi} \in \mc H \bigm|
			Q_B \ket{\psi} = 0, b_0 \ket{\psi} = 0,
			\nexists \ket{\chi} \in \mc H: \ket{\psi} = Q_B \ket{\chi}
			\Big\}.
\end{equation}
Studying this cohomology is very convenient because the following observation allows to simplify the BRST operator when working in this subspace.
The BRST operator \eqref{general_BRST_charge} can be decomposed in terms of the ghost zero-modes as:
\begin{equation}
	\label{brst:eq:splitting-Q}
	Q_B = c_0 L_0 - b_0 M + \what Q
\end{equation}
where
\begin{equation}
	\label{brst:eq:Q-hat}
	\what Q = \sum_{n \neq 0} c_{-n} L^m_n
		- \frac{1}{2} \sum_{\substack{m,n \neq 0 \\ m + n \neq 0}} (m - n)\, c_{-m} c_{-n} b_{m+n}\ , \qquad
	M = \sum_{n \neq 0} n\, c_{-n} c_n.
\end{equation}
Nilpotency of the BRST operator implies the relations:
\begin{equation}
	\com{L_0}{M} = \com{\what Q}{M}
		= \com{\what Q}{L_0}
		= 0,
	\qquad
	\what Q^2 = L_0 M.
\end{equation}
In the subspace $\mc H_0$, the BRST operator reduces then to $\what Q$, which also becomes nilpotent
\begin{equation}
	\ket{\psi} \in \mc H_0
	\quad \Longrightarrow \quad
	Q_B \ket{\psi}
		= \what Q \ket{\psi},
	\qquad
	\what Q^2 \ket{\psi} = 0.
\end{equation}
It is then sufficient to work out the BRST cohomology for the $\what Q$ operator:
\begin{equation}
	\mc H_{\text{rel}}(Q_B, \mc H)
		\simeq \mc H(\what Q, \mc H_0).
\end{equation}
Note though that on-shellness does not imply $b_0 \ket{\psi} = 0$, i.e.\ there are on-shell states that have $b_0 \ket{\psi} \neq 0$.
Hence, once the relative cohomology has been constructed, one needs to build the absolute cohomology by relaxing the condition $b_0 = 0$.

\subsection{Light-cone parametrization}
\label{sec:brst_derivation:lc}

Following the general method, it is useful to introduce a light-cone parametrization for the scalar fields.
We generalise it here to include both the spacelike and timelike field cases by introducing the parameter $\epsilon_\phi$:
\begin{subequations}
\begin{gather}
	\alpha^\pm_n
		= \frac{1}{\sqrt{2}}\, \left(\alpha^\phi_n \pm\frac{\I}{\sqrt{\epsilon_\phi}\,} \alpha^X_n\right),
	\qquad
	\alpha^\pm
		= \frac{1}{\sqrt{2}}\, \left(\epsilon_\phi\, \alpha_\phi \pm \frac{\I}{\sqrt{\epsilon_\phi}}\, \alpha_X\right),
	\\
	x^\pm
		= \frac{1}{\sqrt{2}}\, \left( \epsilon_\phi\,x_\phi \pm \frac{\I}{\sqrt{\epsilon_\phi}} \, x_X\right),
	\qquad
	P^\pm
		= \frac{1}{\sqrt{2}}\, \left(\epsilon_\phi\, P_\phi \pm \frac{\I}{\sqrt{\epsilon_\phi}}\, P_X\right),\label{gen-momenta}
	\\
	Q^\pm
		= \frac{1}{\sqrt{2 }}\, \left(Q_\phi \pm\frac{\I}{\sqrt{\epsilon_\phi}\,} Q_X\right).\label{gen-charges}
\end{gather}
\end{subequations}
Using $ \alpha= \epsilon\, Q/2+\I P$, one obtains:
\begin{equation}
	\alpha^\pm = \frac{ Q^\pm}{2}+\I P^\pm.
\end{equation}
The commutations relations are
\begin{equation}
	\com{\alpha^+_m}{\alpha^-_n}
		= \epsilon_\phi\, m\, \delta_{m+n,0},
	\qquad
	\com{x^\pm}{P^\mp}
		= \I \frac{\epsilon_\phi}{2}.
\end{equation}
The light-cone level and number operators are defined as:
\begin{equation}
	N_\pm = \sum_{n > 0} n\, N^\pm_n, \qquad
	N^\pm_n = \frac{\epsilon_\phi}{n}\, \alpha_{-n}^\pm \alpha_n^\mp,
\end{equation}
such that
\begin{equation}
	N^X + N^\phi
		= N^+ + N^-.
\end{equation}
Using these definitions, the expression \eqref{brst:eq:L0-total} for the total Virasoro zero-mode becomes
\begin{subequations}
\begin{gather}
	\label{brst:eq:on-shell-condition}
	L_0
		= L^\perp_0 - m^2 - 1 + \what L_0^\|, \\
	\label{brst:eq:level-operator-lc}
	\what L_0^\|
		= N^+ + N^- + N^b + N^c, \\
	\label{brst:eq:mass-lc}
	- m^2
		= \frac{ \epsilon_\phi}{2}\, \left( Q^+ Q^- + 4 P^+P^-\right)
		= 1 - \frac{c_\perp}{24} + 2 \epsilon_\phi P^+ P^-.
	\end{gather}
\end{subequations}

We further define the momenta\footnote{In~\cite{Bautista:2019:QuantumGravityTimelike}, the notation for the generalised momenta is $K^\pm:= P^\pm$.}
\begin{equation}\label{brst:eq:P-pm-n}
P^\pm_n
	= - \I \alpha^\pm+ \frac{\I}{2}\, Q^\pm (n+1)
	=  P^\pm + \frac{\I}{2}\, Q^\pm \, n,
\end{equation}
whose zero-modes are related to the cylinder light-cone momenta as
\begin{equation}
P^\pm_0= P^\pm.
\end{equation}

\subsection{Reduced cohomology}
\label{sec:brst_derivation:red}

Introducing the above light-cone parametrization into the $\what Q$ operator allows to further decompose it as
\begin{equation}
	\label{brst:eq:splitting-hat-Q}
	\what Q = Q_0 + Q_1 + Q_2,
\end{equation}
where
\begin{equation}
	\label{brst-der:Q-decomp-012}
	\begin{gathered}
		Q_1 = \sum_{n\neq 0} c_{-n} L^\perp_n
			+ \sum_{\substack{m,n \neq 0 \\ m + n \neq 0}} c_{-m} \left(\epsilon_\phi\,\alpha^+_{-n} \alpha^-_{m+n} - \frac{1}{2}\, (m - n)\, c_{-n} b_{m+n} \right),
		\\
		Q_0 = \epsilon_\phi\sqrt{2}\sum_{m \neq 0} P^+_m\, c_{-m} \alpha^-_m,
		\qquad
		Q_2 = \epsilon_\phi\sqrt{2}\sum_{m \neq 0} P^-_m\, c_{-m} \alpha^+_m.
	\end{gathered}
\end{equation}
The subscripts $0,1,2$ refer to the degree of the operator defined as $N^+ - N^- + N^c - N^b$.
Nilpotency of $\what Q$ gives the following conditions:
\begin{equation}
	\label{brst:eq:conditions-Qj}
	Q_0^2 = Q_2^2 = 0, \qquad
	\anticom{Q_0}{Q_1} = \anticom{Q_1}{Q_2} = 0, \qquad
	Q_1^2 + \anticom{Q_0}{Q_2} = 0.
\end{equation}
Therefore, $Q_0$ and $Q_2$ are both nilpotent and define a cohomology.
The whole point of this decomposition and of the light-cone parametrisation is that the cohomologies of $\what Q$ and $Q_0$ are isomorphic\footnotemark{}
\footnotetext{%
	The role of $Q_0$ and $Q_2$ can be reversed by changing the sign in the definition of the degree and the role of $P^\pm_n$.
}
\begin{equation}
	\mc H(\what Q, \mc H_0) \simeq \mc H(Q_0, \mc H_0)
\end{equation}
when there is at most one degree for each ghost number~\cite{Bouwknegt:1992:BRSTAnalysisPhysical}.
In particular, this holds automatically if there are no ghosts and no light-cone oscillators.
Note that $Q_0$ commutes with $L_0$ and $b_0$: to compute the cohomology of $\what Q$, one can compute the cohomology of $Q_0$ for the Fock space $\mc H$ and restrict it at the end to $\mc H_0$.

The computation of the $Q_0$- and $Q_2$-cohomologies requires to invert either one of the momenta $P_n^\pm$.
A subtlety therefore arises if both $P^\pm_n$ vanish for some integers.
In this case, some oscillators are not present in the expression of the contracting homotopy operator and one finds more states in the cohomology.
For this reason, we will deal separately with the two cases: 1) $P^+_n \neq 0$ for all $n\neq0$  ($P^-_n$ can vanish for some value) and 2) $P^+_r = P^-_s = 0$ for some $r, s \neq 0$.
The case $P^-_n\neq 0$ for all $n\neq0$ ($P^+_n$ can vanish for some value) is analogous to the first case by reversing the definition of the degree.

\paragraph{Non-vanishing $P^\pm_n$: continuous states.}

When $P^+_n \neq 0$ for all $n$, one can introduce the contracting homotopy operator
\begin{equation}
	B = \frac{1}{\sqrt{2}}\sum_{n \neq 0} \frac{1}{P^+_n}\, \alpha^+_{-n} b_n
\end{equation}
such that
\begin{equation}\label{commut}
	\what L_0^\| = \anticom{Q_0}{B}.
\end{equation}
A state $\ket{\psi} \in \mc H$ is in the $Q_0$-cohomology only if
\begin{equation}
	\label{on-shell_reduced}
	\what L_0^\| \ket{\psi} = 0.
\end{equation}
Indeed, following the same reasoning as in \eqref{brst:eq:off-shell-psi}, one finds that
\begin{equation}
	\ket{\psi} = \frac{1}{\what L_0^\|} \, Q_0 \big( B \ket{\psi} \big)
\end{equation}
if $\what L_0^\| \ket{\psi} \neq 0$, then $\ket{\psi}$ is $Q_0$-exact hence not in the cohomology.
Since $\what L_0^\| = N^+ + N^- + N^b + N^c$ is a sum of positive numbers, \eqref{on-shell_reduced} implies that each of them has to vanish.
As a consequence, states in the cohomology do not contain any $\alpha_n^\pm$, $b$ or $c$ excitations.
Since these states are built only from transverse excitations, they cannot be $Q_0$-exact.

The next step is to prove that $\what L_0^\|= 0$  states are closed.
These states have $N_{\text{gh}} = 1$ since they contain only the ghost vacuum $\ket{\downarrow}$ ($N^b = N^c = 0$).
Further, given that $\what L_0^\|$ and $Q_0$ commute, as follows from \eqref{commut}, one has
\begin{equation}
	0 = Q_0 \what L_0^\| \ket{\psi}
		= \what L_0^\| Q_0 \ket{\psi}.
\end{equation}
Since $Q_0$ increases the ghost number of $\ket{\psi}$ by $1$, one can invert $\what L_0^\|$ in the last term since $\what L_0^\|\neq 0$ in this subspace.
This gives
\begin{equation}
	Q_0 \ket{\psi} = 0,
\end{equation}
hence these states are closed as announced.
Hence, $\what L_0 = 0$ is both a necessary and sufficient condition.

Finally, since these states only contain the ghost vacuum, they automatically satisfy $b_0 \ket{\psi}=0$.
The only remaining condition to impose is the on-shell condition $L_0 \ket{\psi} = 0$ with the zero-mode given by \eqref{brst:eq:L0-total-simp}, which on these states becomes:
\begin{equation}
	\label{brst:eq:on-shell-standard-states}
	L_0
		= L^\perp_0 - \frac{c_\perp}{24} + \big( p_X^2 + \epsilon_\phi \, p_\phi^2 \big)
		= 0.
\end{equation}
As a consequence, states in the cohomology do not contain any $\alpha_n^\pm$, $b$ or $c$ excitations: they don't contain ghosts and correspond to the ground state of the Fock space $\mc F_\phi \otimes \mc F_X$ with momenta constrained by the above condition.
The states satisfying this condition will be called continuous states\footnotemark{} since one of the momenta varies continuously.
\footnotetext{%
	This term is used in opposition to “discrete states”, to be defined shortly.
	However, the momentum may not be continuous if the Coulomb gas $X$ is reduced to minimal models and if the transverse CFT has a discrete spectrum.
	In~\cite{Bautista:2019:QuantumGravityTimelike}, these states were named standard.
}%
Since they all have the same degree, one can apply the theorem A.3 from~\cite{Bouwknegt:1992:BRSTAnalysisPhysical} to show that they are also elements of the $\what Q$-cohomology.\footnotemark{}
\footnotetext{%
	The theorem states that the $Q_0$- and $\what Q$-cohomologies are isomorphic if states at fixed ghost number have all the same degree.
}%

\paragraph{Vanishing $P^\pm_n$: discrete states.}

If there exist two non-zero integers $r$ and $s$ such that the operators $P^\pm_n$ vanish
\begin{equation}
	\label{brst:eq:vanishing-Pr-Ps}
	\exists\quad r, s \in \Z^*:
	\qquad
	P^+_r = P^-_s
		= 0,
\end{equation}
one can instead introduce a modified contracting homotopy operator:
\begin{equation}
	B_r
		= \frac{1}{\sqrt{2}}\sum_{n \neq 0, r} \frac{1}{P^+_n}\, \alpha^+_{-n} b_n
\end{equation}
such that
\begin{equation}
	\what L_{0,r}^\| = \anticom{Q_0}{B_r}.
\end{equation}
By the same argument as in the previous case, a state $\ket{\psi}$ is in the cohomology only if
\begin{equation}
	\what L_{0,r}^\| \ket{\psi} = 0.
\end{equation}

Given that
\begin{equation}
	\what L_0^\| - \what L_{0,r}^\|
		= \anticom{Q_0}{B - B_r}
		= \epsilon_\phi\anticom{c_{-r} \alpha^-_r}{\alpha^+_{-r} b_r},
\end{equation}
the level operator differs from the modified one by
\begin{align}
	\label{brst:eq:discrete-states-hat-L}
	r > 0 &:
	\qquad
	\what L_0^\| = \what L_{0,r}^\| + r \, \big( N^+_r + N^c_r \big) \\
	r < 0 &:
	\qquad
	\what L_0^\| = \what L_{0,r}^\| - r \, \big( N^-_{-r} + N^b_{-r} \big)
\end{align}
since $b_r$ ($c_r$) is an annihilation operator associated to $c_{-r}$ ($b_{-r}$) and $\alpha^+_{-r}$ ($\alpha^-_{-r}$) is a creation operator associated to $\alpha^-_r$ ($\alpha^+_r$) for $r > 0$ ($r < 0$).
States in the cohomology can hence be built by acting with the corresponding creation operators on the vacuum and are called discrete states:
\begin{subequations}
\label{brst:eq:discrete-states-ghost}
\begin{align}
	r > 0 &:
	\qquad
		(\alpha^+_{-r})^u \, (c_{-r})^v \ket{p_\phi, p_X, \downarrow}\otimes \ket{\psi_\perp},
		\\
	r < 0 &:
	\qquad
		(\alpha^-_r)^u \, (b_r)^v \ket{p_\phi, p_X, \downarrow}\otimes \ket{\psi_\perp},
\end{align}
\end{subequations}
where $u$ and $v$ are some positive integers to be determined in each case by consistency with the other conditions.
The allowed values for $v$ are $0$ and $1$: the cohomology will hence contain states with ghost number $N_{\text{gh}} = 0, 1, 2$.

We now impose the on-shell condition.
The vanishing of the two momenta \eqref{brst:eq:vanishing-Pr-Ps}, can be used to recast the expression of the momenta \eqref{brst:eq:P-pm-n} as:
\begin{equation}
	\label{brst:eq:P-pm-n-discr}
	P^+_m
		= \frac{\I\, Q^+}{2} (m - r),
	\qquad
	P^-_m
		= \frac{\I\, Q^-}{2} (m - s).
\end{equation}
In particular, one finds
\begin{equation}
	\label{brst:eq:momenta-prod}
	P^+ P^-
		= -\epsilon_\phi\, \frac{r s}{2}  \, \left(1 - \frac{c_\perp}{24} \right).
\end{equation}
These states will be called discrete states because their momenta always take on discrete values, as follows from \eqref{brst:eq:P-pm-n-discr}.
The on-shell condition \eqref{brst:eq:on-shell-condition} becomes
\begin{equation}
	\label{brst:eq:on-shell-discrete-states}
	L_0
		= L^\perp_0+ (1 - r s) \, \left(1 - \frac{c_\perp}{24} \right)-1 + \what L_{0}^\|
		= 0.
\end{equation}
If there is no transverse CFT, this equation can only have a solution if $r s > 0$.
On the other hand, this equation may admit solutions with $r s < 0$ if there is a transverse CFT.\footnotemark{}
\footnotetext{%
	We have ignored the subtleties at zero-momentum since that requires to know the transverse CFT.
	Indeed, if $p_\phi = p_X = 0$ but there is another scalar field in the transverse CFT, then it is possible to use that one instead of $X$.
}%
Applying the on-shell condition \eqref{brst:eq:on-shell-discrete-states} on the states \eqref{brst:eq:discrete-states-ghost} yields
\begin{equation}
	\label{brst:eq:on-shell-discrete-states-ab}
	L_0
		= L^\perp_0+ (1 - r s) \, \left(1 - \frac{c_\perp}{24} \right) - 1 + \abs{r}(u + v)
		= 0.
\end{equation}

For fixed momenta of the vacuum $p_X$ and $ p_\phi$, the values of the indices $r$ and $s$ are determined from the vanishing of $P_r^+$ and $P_s^-$ (given that the charges $Q_\phi$ and $Q_X$ are fixed by the theory).
Hence, once the momenta are fixed, only one case between $r > 0$ or $r < 0$ can take place: states in the cohomology of a fixed-momentum sector differ by at most one ghost number (i.e.\ $N_{\text{gh}} = 1,2$ or $N_{\text{gh}} = 0, 1$).

It is not possible to apply theorem A.3 from~\cite{Bouwknegt:1992:BRSTAnalysisPhysical} to show that the discrete states \eqref{brst:eq:discrete-states-ghost} are also elements of the $\what Q$-cohomology.
Indeed, there can be states of different degrees with $N_{\text{gh}} = 1$ satisfying the on-shell condition \eqref{brst:eq:on-shell-discrete-states-ab}.
One such example is the case where $Q_X = 0$ and the transverse CFT is made of $D - 1$ scalars~\cite{Bilal:1992:RemarksBRSTcohomologycM}.

On the other hand, discrete states contain ghost excitations and can have negative-norm states.
For these reasons, it is useful to project them out, which can be achieved by imposing Hermiticity of the BRST charge.

\paragraph{Hermiticity and absolute cohomology.}

Hermiticity of the BRST charge follows from the Hermiticity of the Virasoro operators of each sector.
For the Coulomb gas, the standard Hermiticity conditions  \eqref{cft:eq:hermiticity-coulomb} require $P^\dagger = P$ if $Q \in \R$, and $P^\dagger = - P$ if $Q \in \I \R$.

For continuous states, this simply restricts the range of the momenta $p_X$ and $p_\phi$.
Note that this is independent of $\epsilon_\phi = \pm 1$.

To study the discrete states, we first rewrite the momenta for the scalar fields when $P^+_r = P^-_s = 0$ hold:
\begin{subequations}
\label{brst:eq:discrete-states-momenta-bis}
\begin{align}
	P_\phi
		&= - \frac{\I \epsilon_\phi}{4} (r + s) Q_\phi + \frac{\sqrt{\epsilon_\phi}}{4}\, (r - s) Q_X ,
	\\
	P_X
		&= - \frac{\sqrt{\epsilon_\phi}}{4} (r - s) Q_\phi - \frac{\I}{4} (r +s) Q_X,
\end{align}
\end{subequations}
using \eqref{brst:eq:P-pm-n} and \eqref{gen-momenta}, \eqref{gen-charges}.
As a consequence, restricting the BRST cohomology to its Hermitian sector imposes constraints on the possible values of $r$ and $s$.
Tables~\ref{table-ghost-states-timelike} and \ref{table-ghost-states-spacelike} display the conditions for which $X$ is Hermitian for the timelike and spacelike Liouville cases respectively:
\begin{description}
	\item[Timelike case]
		We find that the only allowed possibility is for $r = s = 0$ when both fields are Hermitian, and hence there are no discrete states.

	\item[Spacelike case]
		When both fields are Hermitian, the only solution is $r = s = 0$ if one charge is real and the other imaginary.
		On the other hand, there are solutions with $r = - s$ when the charges are either both real or both imaginary.
		If there is no transverse CFT, then $r = - s$ does not solve the on-shell condition \eqref{brst:eq:on-shell-discrete-states-ab} since $r s < 0$.\footnotemark{}
		\footnotetext{%
			If there is a transverse CFT, then $c_\perp < 24$ and it looks difficult for solutions to \eqref{brst:eq:on-shell-discrete-states-ab} to exist when both charges are real.
			More generally, if the transverse CFT contains a scalar field (with or without background charge), it can be used instead of $X$.
			In that case, one may reduce the computations of the cohomology to another case of the table and find that, in fact, there are no discrete states.
	}%
		We leave open the case where there is a transverse CFT.
\end{description}

\begin{table}[htp]
\begin{center}
\begin{tabu}{c| [1pt] c|c}
	$\epsilon_\phi=-1$ & $Q_X \in \R$ & $Q_X \in \I\R$ \\ \tabucline[1pt]{-}
	$Q_\phi \in \R$ & $r=s=0$ & $r+s=0$ \\ \hline
	$Q_\phi \in \I \R$ &$r+s=0$ & $r=s=0$
\end{tabu}
\caption{
	Conditions on the integers $r$ and $s$ for a \emph{timelike} $\phi$ following from the Hermiticity of $X$.
	Imposing further $\phi$ to be Hermitian reduces all conditions to $r = s = 0$, in which case there are no solutions.
}
\label{table-ghost-states-timelike}
\end{center}
\end{table}

\begin{table}[htp]
\begin{center}
\begin{tabu}{c| [1pt] c|c}
	$\epsilon_\phi=+1$& $Q_X \in \R$ & $Q_X \in \I\R$ \\ \tabucline[1pt]{-}
	$Q_\phi \in \R$ & $r+s=0$& $r=s=0$ \\ \hline
	$Q_\phi \in \I \R$ & $r=s=0$ & $r+s=0$
\end{tabu}
\caption{
	Conditions on the integers $r$ and $s$ for a \emph{spacelike} $\phi$ following from the Hermiticity of $X$.
	Hermiticity of $\phi$ does not impose further conditions.
}
\label{table-ghost-states-spacelike}
\end{center}
\end{table}

In conclusion, there are no discrete states, hence no ghosts, in the Hermitian subsector of the BRST cohomology in most cases.
Indeed, in the case the Liouville field (or one of the two Coulomb gases) is timelike, only continuous states remain.
In that case, it is possible to apply theorem A.3 from~\cite{Bouwknegt:1992:BRSTAnalysisPhysical} which implies that the $Q_0$-cohomology is isomorphic to the relative cohomology.
Finally, the absolute cohomology follows trivially as $\mc H_{\text{abs}} = \mc H_{\text{rel}} \oplus c_0 \mc H_{\text{rel}}$~\cite{Bouwknegt:1992:BRSTAnalysisPhysical}.
It should be noted that the Hermiticity conditions preventing discrete states to appear in the spectrum are those of the two Coulomb gas fields $X$ and $\phi$.
However, we assume Hermiticity of the whole matter sector so that the BRST charge is Hermitian.

\paragraph{Cosmological constant.}

Finally, we outline a simple argument to argue that restoring the cosmological constant does not change the cohomology beyond identifying states with $ p$ and $-p$.
This relies on the isomorphism between the Fock basis $\{ \alpha_{-n} \}$ and the Virasoro basis $\{ L_{-n} \}$, which holds when the momentum of the vacuum $\ket{p}$ \eqref{cft:def:vacua} is not equal to the momentum \eqref{eq:degen-momenta} of a degenerate state, $p \neq p _{r,s}$~\cite{Iohara:2013:RepresentationTheoryVirasoro, Schottenloher:2008:MathematicalIntroductionConformal, Ribault:2016:FreeBosonsVirasoro, Frenkel:1986:SemiinfiniteCohomologyString}.\footnotemark{}
\footnotetext{%
	A hint of the proof is the following.
	Both bases have the same number of states at a given level because $\alpha_{-n}$ can be replaced with $L_{-n}$, or the other way around.
	In particular, both characters are proportional to $\eta(\tau)^{-1}$.
	There is a map from the Virasoro basis to the Fock basis using \eqref{eq:cg-Ln}.
	It remains to show that the map is one-to-one.
	The Fock basis is obviously not degenerate, so the same must hold for the Virasoro basis, which can be checked using the Kac determinant.
}%
This is mostly the case since, as discussed below \eqref{cft:eq:hermiticity-coulomb}, there are no degenerate states for Hermitian momenta when $c_L \notin (1, 25)$ for both $\epsilon_\phi = \pm 1$.
The argument is as follows:
\begin{enumerate}
	\item The cohomology is computed for the Coulomb gas in the oscillator basis $\{ \alpha_{-n} \}$, that is, as a subspace of the Fock space built on all primaries $V_{\pm p}$.

	\item The isomorphism is used to rewrite \emph{all states} in the Virasoro basis $\{ L_{-n} \}$, which is possible when there are no degenerate states.
	In particular, we map the states from the cohomology.

	\item The Liouville states are written as linear combinations \eqref{reflected-op} of Coulomb gas states in the Virasoro basis.
	This can be done because Liouville primaries are given by
	\begin{equation}\label{linear-combi}
		 \mc V_p
			= V_p + R(p) V_{-p}.
	\end{equation}
	Therefore their descendents are generated by the Virasoro basis $\{ L_{-n} \}$ acting on these.
	Since they are written as linear combinations of states $V_{\pm p}$,  which have identical conformal weights $h_p$, they also have a well-defined conformal weight (and similarly for the descendents).

	\item The BRST charge $Q_B$ is completely determined by the Virasoro algebra, and its result depends only on the conformal weight of the states and the central charge $c_L$.
	Since the central charges and conformal weights of the Liouville theory and of the corresponding Coulomb gas are the same, the Liouville cohomology follows from the one of the Coulomb gas.
\end{enumerate}
Notice that it follows from \eqref{brst:eq:on-shell-standard-states} that if a \emph{continuous} state $V_p$ is in the cohomology, so is $V_{-p}$.
(This is not the case though for discrete states, so the cosmological constant could have a different effect for these states.)
In particular, this works for both the spacelike and timelike Liouville theories at $c_L \le 1$ and $c_L \ge 25$ when restricting to the Hermitian subspace.\footnotemark{}
\footnotetext{%
	For this approach to work, it is necessary to restrict the states to Hermitian momenta from the outset, since the isomorphism (which relies on Hermiticity) must be used before computing the Liouville cohomology.
}%

For another approach which could be extended to more general cases, the reader is referred to~\cite{Bautista:2019:QuantumGravityTimelike}.

%% file: sections/brst_applications.tex
\section{Applications}
\label{sec:brst_applications}

In this section, we apply the results from \Cref{sec:brst_derivation} to the three Liouville theories which have a real action (at least for the Gaussian or Coulomb gas part): spacelike with $c_L \ge 25$, with $c_L \in (1, 25)$, and timelike with $c \le 1$.
Due to the reality of the action, these models are the most natural ones for defining a theory of $2d$ gravity.
In each case, we consider the coupling to different simple models of spacelike matter.

The relevant formulas to solve for each theory are the relation \eqref{brst:eq:square-charges} between the Coulomb charges and the central charge of the transverse CFT, the on-shell conditions \eqref{brst:eq:on-shell-standard-states} and \eqref{brst:eq:on-shell-discrete-states-ab}, and the expressions for the momenta in the presence of discrete states \eqref{brst:eq:discrete-states-momenta-bis}.

\subsection{Timelike Liouville with \texorpdfstring{$c_L \le 1$}{cL <= 1}}
\label{sec:brst-timelike}

We consider the timelike Liouville theory $\epsilon_\phi = - 1$ with $c_L \le 1$, that is $Q_\phi \in \R$.
Since the charge is real, the Hermiticity condition implies that the Liouville momentum is real, $p_\phi \in \R$.

\paragraph{Free scalar fields}

We consider $D \ge 25$ scalar fields $X^I = (X, X^i)$ with $i = 1, \ldots, D - 1$ and such that $c_\perp = D - 1$, $Q_X = 0$~\cite{Bautista:2019:QuantumGravityTimelike}.\footnote{In~\cite{Bautista:2019:QuantumGravityTimelike}, the notation used is $Y^i:= X^i $ for the transverse scalars,  the Liouville field is denoted $\chi$ instead of $\phi$, and the momenta are $k^i:= p^i$, $K:= p_X$ and $E:= p_\phi$.}
The momenta are denoted as $\vec p = (p_X, p^i)$ and the transverse level operator by $N_\perp \in \N$.
The background charge is related to $D$ as
\begin{equation}
	Q_\phi^2 = \frac{D - 25}{6}.
\end{equation}
The on-shell condition reads
\begin{equation}
	{\vec p}^2 - p_\phi^2 + N_\perp = \frac{D - 1}{24}.
\end{equation}
This equation is solved as
\begin{equation}
	p_\phi = \pm \sqrt{{\vec p}^2 + N_\perp - \frac{D - 1}{24}}.
\end{equation}
According to the Hermiticity conditions, $p_\phi, \vec p \in \R$, which puts a lower bound on $\abs{\vec p}$ for the square root to have a positive argument.
As noted previously, there are no ghost states and this reproduces the results from~\cite{Bautista:2019:QuantumGravityTimelike} when those are restricted to the Hermitian sector.
For $Q_\phi = 0$ one recovers string theory in $D = 26$ dimensions.

\paragraph{Real Coulomb gas}

Consider the coupling of Liouville to a single Coulomb gas.
In this case, the vanishing of the total central charge gives the relation
\begin{equation}
	Q_X^2 - Q_\phi^2 = 4,
\end{equation}
which implies that $Q_X \in \R$.
The solutions to the on-shell condition are
\begin{equation}
	p_\phi = \pm p_X \in \R
\end{equation}
which is compatible with Hermiticity.

\paragraph{Real Coulomb gas}

We recall the relation \eqref{brst:eq:square-charges} between the Coulomb charges and the central charge of the transverse CFT
\begin{equation}
	Q_X^2 + Q_\phi^2
		= 4 - \frac{c_\perp}{6}.
\end{equation}
In order to describe the continuous states, one just needs to solve the on-shell condition \eqref{brst:eq:on-shell-standard-states}
\begin{equation}
	L_0
		= L_0^\perp - \frac{c_\perp}{24} + p_X^2 + p_\phi^2
		= 0.
\end{equation}
In this case, there are no ghost states and no light-cone oscillators.

In the spacelike case, there can be discrete states \eqref{brst:eq:discrete-states-ghost}.
The corresponding on-shell equation \eqref{brst:eq:on-shell-discrete-states-ab} reads:
\begin{equation}
	L_0
		= L^\perp_0 + (1 - r s) \, \left(1 - \frac{c_\perp}{24} \right) - 1 + \abs{r}(u + v)
		= 0.
\end{equation}
where $r$ and $s$ are such that $P_r^+ = P_s^- = 0$.
Then, the momenta of the discrete states are given by \eqref{brst:eq:discrete-states-momenta-bis}:
\begin{equation}
	\begin{aligned}
	P_\phi
		&
		= - \frac{\I \epsilon_\phi}{4} \, (r + s) Q_\phi + \frac{\sqrt{\epsilon_\phi}}{4} \, (r - s) Q_X,
	\\
	P_X
		&
		= - \frac{\sqrt{\epsilon_\phi}}{4} (r - s) Q_\phi - \frac{\I}{4} (r +s) Q_X.
	\end{aligned}
\end{equation}

\subsection{Spacelike Liouville with \texorpdfstring{$c_L \ge 25$}{cL >= 25}}

For the spacelike Liouville theory $\epsilon_\phi = 1$ with $c_L \ge 25$, $Q_\phi$ is real and $Q_\phi \ge 2$.
Since the background charge is real, the Hermiticity condition implies that the Liouville momentum is real, $p_\phi \in \R$.

We consider two matter models without transverse CFT such that
\begin{equation}
	Q_\phi^2 + Q_X^2 = 4.
\end{equation}
Since $Q_\phi \ge 2$, this admits a solution only if $Q_X = 0$ or $Q_X \in \I \R$.

According to the discussion in \Cref{sec:brst_derivation:red}, the on-shell condition for discrete states admits no solutions if there is no transverse CFT, since in this case $r s \le 0$.
As a consequence, there are no ghost states in the Hermitian subsector.

\paragraph{Free scalar field}

In this case, $X$ is a free scalar field CFT with $c_X = 1$ and $Q_X = 0$~\cites[sec.~5]{Bouwknegt:1992:BRSTAnalysisPhysical}{Lian:1991:2DGravityC1, Mukherji:1991:NullVectorsExtra}, which implies $Q_\phi = 2$ and $c_L = 25$.
The on-shell equation for continuous states reads
\begin{equation}
	p_X^2 + p_\phi^2 = 0
\end{equation}
whose Hermitian solutions are:
\begin{equation}
	p_\phi \in \R,
	\qquad
	p_X = \pm \I p_\phi \in \I \R.
\end{equation}

\paragraph{Imaginary Coulomb gas}

We consider the case where $X$ is a Coulomb gas with imaginary charge $Q_X \in \I \R$ such that $c_X \le 1$.
The on-shell equation is the same as for the free scalar and the Hermitian solutions are:
\begin{equation}
	p_\phi \in \R,
	\quad
	p_X = \pm \I p_\phi \in \I \R.
\end{equation}

In this case of an imaginary Coulomb gas, $c_X$ matches the central charge of a minimal model at some specific values.
Indeed, minimal models $M_{p,q}$ are obtained when $b^2 = - q/p$, with $p, q \ge 2$, is a rational number such that
\begin{equation}
	c_m = 1 - \frac{6 (p - q)^2}{p q} < 1.
\end{equation}
Minimal models can be defined from a Coulomb gas with imaginary background charge thanks to Felder's resolution~\cite{Felder:1989:BRSTApproachMinimal, Bouwknegt:1991:FockSpaceResolutions} (see also~\cite{Bouwknegt:1992:BRSTAnalysisPhysical}) to obtain irreducible representations; this is well known and we refer the reader to the literature~\cite{Bouwknegt:1992:BRSTAnalysisPhysical, Lian:1991:NewSelectionRules, Mukhi:1991:ExtraStatesC1, Imbimbo:1992:ConstructionPhysicalStates} for more details.

We expect this to be also true for generalized minimal models~\cite{Zamolodchikov:2005:ThreepointFunctionMinimal, Ribault:2014:ConformalFieldTheory, Ribault:2015:LiouvilleTheoryCentral}.
These models exist for any complex central charge $c \in \C$, but only on the sphere.
Coupling them to gravity gives the so-called “generalized minimal gravity”~\cite{Zamolodchikov:2005:ThreepointFunctionMinimal} which allows to extend minimal gravity defined at discrete points in $c_L \in [25, \infty)$ to all central charges $c_L > 25$ (in fact, also to $c_L \le 25$).
The first step would be to generalize Felder's resolution to describe these models from the Coulomb gas.

\subsection{Spacelike Liouville with \texorpdfstring{$c_L \in (1, 25)$}{cL = (1, 25)}}

For the spacelike Liouville theory $\epsilon_\phi = 1$ with $c_L \in (1, 25)$, $Q_\phi$ is real and $Q_\phi \in (0, 2)$
Since the background charge is real, the Hermiticity condition implies that the Liouville momentum is real, $p_\phi \in \R$.
The BRST cohomology of this theory has been studied in~\cite{Bilal:1992:RemarksBRSTcohomologycM}, but with a different Hermiticity condition.\footnotemark{}
\footnotetext{%
	In~\cite{Bilal:1992:RemarksBRSTcohomologycM}, the oscillator zero-mode receives an additional shift (for $\epsilon = 1$): $L_n(\alpha)^\dagger = L_{-n}( \epsilon Q - \alpha)$.
	This leads to the same conditions as in \eqref{cft:eq:hermiticity-coulomb} except for $P^\dagger = \mp P$.
	However, this implies that the momentum is anti-Hermitian in the limit $Q \in \R \to 0$.
}%

\paragraph{Real Coulomb gas}

We consider the case where there is no transverse CFT.
The equation for the background charge becomes
\begin{equation}
	Q_X^2 + Q_\phi^2 = 4,
\end{equation}
which implies that $Q_X \in (0, 2) \in \R$.
In this case, the on-shell condition for continuous states reads:
\begin{equation}
	p_X^2 + p_\phi^2
		= 0.
\end{equation}
The only solution in the Hermitian sector $p_X, p_\phi \in \R$ corresponds to $p_X = p_\phi = 0$.
This means that the only state in the cohomology is the vacuum.

Hermiticity allows discrete states with $r = - s$, however, the on-shell condition does not have any solution since $r s < 0$.

\paragraph{Free scalar fields}

We consider $D \in (1, 25)$ scalar fields $X^I = (X, X^i)$ with $i = 1, \ldots, D - 1$ and such that $c_\perp = D - 1$, $Q_X = 0$ (if $Q_X \neq 0$, then one can perform a rotation in the field space to recover this case~\cite{Bilal:1992:RemarksBRSTcohomologycM}).
The momenta are denoted as $\vec p = (p_X, p^i)$ and the transverse level operator by $N_\perp \in \N$.
The background charge reads
\begin{equation}
	Q_\phi^2 = \frac{25 - D}{6}.
\end{equation}
The on-shell equation is
\begin{equation}
	\vec p^2 + p_\phi^2 + N_\perp
		= \frac{D - 1}{24}.
\end{equation}
Since $D > 1$, the RHS is positive and the solutions are
\begin{equation}
	p_\phi = \pm \sqrt{\frac{D - 1}{24} - \vec p^2 - N_\perp} \,,
\end{equation}
with the values of $\abs{\vec p}$ and $N_\perp$ bounded from above such that the square root is positive and $p_\phi \in \R$ consistent with the Hermiticity condition.

The on-shell condition for the discrete states is:
\begin{equation}
	(p^i)^2 + N_\perp + (1 + r^2) \, \left(\frac{25 - D}{24} \right) - 1 + \abs{r}(u + v)
		= 0.
\end{equation}
This admits non-trivial solutions and thus this model can contain ghost states.
The momenta of the discrete states are given by \eqref{brst:eq:discrete-states-momenta-bis}:
\begin{equation}
	p_\phi = 0
	\qquad
	p_X
		= - \frac{r}{2} \, Q_\phi .
\end{equation}

%% file: sections/discussion.tex
\section{Discussion}
\label{sec:discussion}

In this paper, we have analyzed the BRST cohomology of all possible models where the matter CFT contains at least two Coulomb gases, one which is possibly timelike.
This provides a solid basis to study generic matter models coupled to the timelike Liouville theory with $c_L \le 1$ defined in~\cite{Bautista:2019:QuantumGravityTimelike}.
The main result of our analysis is that its spectrum as determined from the Hermitian sector of the BRST cohomology associated to 2d diffeomorphisms, is free from negative-norm states.
When states with Hermitian momenta cannot be degenerate, we provided an argument to extend the cohomology to include the effect of the cosmological constant.

Another conclusion of our analysis is that two theories with the same value of the central charge can have different BRST spectra.
Indeed, consider the Liouville theories with $(\epsilon_\phi = -1, Q_\phi \in \mathbb{R})$ and with $(\epsilon_\phi = +1, Q_\phi \in \I \mathbb{R})$, both coupled to the same spacelike Coulomb gas with $Q_X \in \I \mathbb{R}$ and the same transverse CFT.
The two have the same central charge $c_L \le 1$, yet the first one only has continuous states in the spectrum, while the second one will generically also contain discrete states.
This should not come as a surprise: as it is well known in string theory, the cohomology is empty in Euclidean signature but not in Lorentzian signature.
This highlights the convenience of using two parameters $(\epsilon,Q)$ to distinguish between the four possible regimes of Liouville theory, as following from the choices of the sign of the kinetic term on the one hand and the reality of the background charge on the other.
Additionally, and as explained in the introduction, one same theory can exhibit different spectra depending on the quantization procedure.

The next step would be to define correlation functions in each case, as was done for the $c_L \le 1$ timelike Liouville theory~\cite{Bautista:2019:QuantumGravityTimelike}.
Following~\cite{Bautista:2019:QuantumGravityTimelike}, we expect that correlation functions of a more general timelike theory must be defined by analytic continuation of the external states of correlation functions of the spacelike theory with the \emph{same} central charge.
This contrasts with the folk lore that the timelike $c_L \le 1$ theory is obtained from the spacelike $c_L \ge 25$ theory by analytic continuation of the Coulomb charge.
Keeping the central charge fixed is a natural generalization of the Wick rotation following string field theory~\cite{Pius:2016:CutkoskyRulesSuperstring, Pius:2018:UnitarityBoxDiagram, deLacroix:2017:ClosedSuperstringField, deLacroix:2019:AnalyticityCrossingSymmetry}.
The motivation for this definition is that correlation functions are \emph{not} analytic in the central charge~\cite{Ribault:2014:ConformalFieldTheory, Ribault:2015:LiouvilleTheoryCentral}.
Consider the $4$-point function for definiteness: using factorization, it can be written as an integral of two $3$-point functions connected by the conformal block.
As one analytically continues the external states, the poles of the integrand move and the integration contour must be deformed to avoid them.
An important simplification happens for $c_L \le 1$ since the poles of the integrand don't move due to the properties of the $c_L \le 1$ structure constant~\cite{Bautista:2019:QuantumGravityTimelike}.
However, the poles do move for $c_L \notin (-\infty, 1]$, which ultimately motivates the general prescription proposed in~\cite{Bautista:2019:QuantumGravityTimelike}.

Finally, it would be interesting to study the discrete states for their own sake as they possess a rich mathematical structure in the spacelike $c_L \le 25$ theory~\cite{Lian:1991:2DGravityC1, Imbimbo:1992:ConstructionPhysicalStates, Witten:1992:GroundRingTwo, Kutasov:1992:GroundRingsTheir, Witten:1992:AlgebraicStructuresDifferential, Seiberg:2004:BranesRingsMatrix}.